# $^{29}$Si Nuclear-Spin Decoherence Process Directly Observed by Multiple Spin-Echoes for Pure and Carrier-less Silicon

Shinji Watanabe [1] and Susumu Sasaki [2,3]

[1]*Graduate School of Science and Technology, Niigata University, Niigata 950-2181, Japan*

[2]*Department of Materials Science and Technology, Faculty of Engineering, Niigata University, Niigata 950-2181, Japan*

[3]*Core Research for Evolutional Science and Technology, Japan Science and Technology Corporation (CREST-JST), Japan*

We report on the $^{29}$Si nuclear spin decoherence time at room temperature for a pure (99.99999%, 7N) and carrier-less (p-type, doping level of $10^{15}$cm$^{-3}$) silicon in which $^{29}$Si nuclei are naturally abundant (4.7%). Despite the experimental difficulties coming from the extremely long spin-lattice relaxation time $T_1$ (of the order of $10^4$ s), we have successfully observed a series of spin-echoes of which the time dependence is characterized by the decoherence time $T_2$. We found that the decoherence process deviates from a single Gaussian or Lorentzian function but is well-reproduced by a bi-exponential function with the shorter $T_2^S = 15\pm5$ ms and the longer $T_2^L = 200\pm20$ ms.

Recently, the study of quantum computing using nuclear magnetic resonance (NMR) has been drastically developing,[1-2] because the interaction between the nuclear spins as quantum bits and the environment is much weaker than any other method. In NMR quantum computing, radio-frequency (RF) pulses must be irradiated for manipulating the quantum bits and reading out the result. Thus, the phase decoherence time $T_2$ of the material used as a "CPU" must be longer than the irradiation time of the RF pulses. From this reason, we have been planning to use pure and carrier-less silicon[3] in which the $T_2$ is expected to be sufficiently long. To our knowledge, however, there has been no experimental report on precise $T_2$ values for this material. Very recently, on the other hand, study on line shapes and $T_1$ was reported by Anne and her colleagues[4]. In their study, they observed the Pake's doublet in the line shape of the "29 silicon" sample in which $^{29}$Si nuclei are enriched to 96.9%. In addition, they found that the $T_1$ is more than 2 hours for their pure and carrier-less (n-type, doping level less than $4 \times 10^{15}$ cm$^{-3}$) silicon samples. It is well-known that in pulsed NMR, once an RF pulse is irradiated, it takes more than several times of $T_1$ for the nuclear spins to be reset to the equilibrium state.[5] Thus, in the case of pure and carrier-less silicon, one must wait more than 10 hours to reset the nuclear spins. On the other hand, in the usual spin-echo method to obtain precise $T_2$ values, the nuclear spins must be reset to the equilibrium state whenever the spin-echo intensity is needed. Thus, with the usual method, we are forced to assume that the sensitivity of the measurement system should be precisely stable during the waiting time of more than 10 hours, which is clearly not appropriate in reality. The waiting-time problem originating from the extremely long $T_1$ has made it next to impossible to perform any kind of pulsed NMR experiments for pure and carrier-less silicon. The problem can be solved only by employing a pulse sequence which does not require the procedure of resetting the nuclear spins.

Very recently, using one of these pulse sequences called the "alternating phase Carr-Purcell (APCP)" method, we have succeeded in observing a series of spin-echoes of which the time dependence is characterized by $T_2$. For pure and carrier-less silicon in which $^{29}$Si nuclei are naturally abundant ("natural silicon"), this is the first report on precise measurement of $T_2$ at room temperature to our knowledge. We found that the decoherence process deviates from a single Gaussian or Lorentzian function but is well-reproduced by a bi-exponential function with the shorter $T_2^S = 15 \pm 5$ ms and the longer $T_2^L = 200 \pm 20$ ms. These $T_2$ values experimentally prove that the pure and carrier-less "natural silicon" is a promising material for nuclear-spin computing.

The silicon sample used in the present study is p-type with doping level $1 \times 10^{15}$

cm$^{-3}$, and the purity is 99.99999% (7 N). In this sample, $^{29}$Si with nuclear spin 1/2 is naturally abundant, i.e., 4.7%. Since the sample weight is about 200mg, the number of $^{29}$Si nuclear spins amounts to be of the order of $10^{20}$.

Our equipment consists of a commercially available 6.8-T NMR magnet (Oxford 300/89) and a home-build phase-coherent type NMR spectrometer. The overall measurement system yields the sensitivity of 0.1 $\mu$V, the resolution of 0.2 ppm and the stability of 0.1 ppm/day. To improve the signal-to-noise ratio in the spectrometer, we carefully designed the resonance circuit to be 50-ohm impedance matched. Moreover, we employed the quadrature detection method in addition to two-phase cycling technique. All the measurements were done at room temperature.

Before measuring the $T_2$ process, we optimized the width of 90-degree RF pulse as 20 $\mu$s, and obtained $^{29}$Si-NMR spectrum by fast-Fourier transformation (FFT) of the free-induction decay (FID). With the pulse width of 20 $\mu$s, we obtained the dependence of the spectrum intensity on the waiting time. From the dependence, we found that the $T_1$ is of the order of $10^4$ s, which is extremely long.

To observe multiple spin-echoes without resetting the nuclear spins, one of the simplest ways would be to employ the Carr-Purcell pulse sequence[7] such that

90 (*x*)- $\tau$ - [180 (*x*)- $\tau$ -ECHO- $\tau$ ] (repeat *N* times),

where *x* stands for applying the RF pulse along the *x* axis in the usual "rotating frame"[8]. In this method, however, slight error in the width of the 180-degree RF pulse increases as the number of the input pulses is increased, which results in a misled spin-echo intensity[6]. The slight error in the pulse width is inevitable, because, as a result of the finite length of the sample coil, the strength of the RF magnetic field depends on the distance from the axis of the coil. Thus, instead of the Carr-Purcell sequence, we employed the alternating-phase Carr-Purcell (APCP) pulse sequence, which is described as

90 (*x*)- $\tau$ - [180 (*x*)- $\tau$ -ECHO (odd)- $\tau$ -180 (-*x*)- $\tau$ -ECHO (even)- $\tau$ ](repeat *N* times) .

In this sequence, the nuclear spins that form "echo (even)" exist exactly in the *xy*-plane, whereas the spins that form "echo (odd)" slightly deviate from the *xy*-plane. As a result, the time dependence of the "echo (even)" intensity gives the $T_2$ process. In our case, we optimized the $\tau$ as 3 ms. Note that the two-phase cycling technique works in the phase inversion (i.e., the alternation of +*x* and –*x*) of the APCP.

For the measurements above-mentioned, we found that, due to the waiting-time problem originating from the extremely long $T_1$, uncontrolled room temperature resulted in the fluctuation of the signal intensity even under the same measurement conditions. To solve this problem, we transferred our whole measurement system to an air-conditioned room where the room temperature was controlled within the error of 0.5 K. This helped us to obtain reproducible and reliable data, as expected.

Usually, solid-state NMR experiments are performed on a powdered sample rather than a single crystal, so that the RF field can irradiate the sample sufficiently without suffering from the skin effect due to the electric conductivity. However, for our carrier-less sample, it is expected that the signal intensity should not be reduced, since the skin depth is estimated to be 6mm which is the same size of the sample. Indeed, we found no appreciable difference between the signal intensity of a powdered sample and that of a single crystal. This means that we can safely say the sample used is "carrier-less".

Figure 1 shows that the spectrum consists of a main peak and two satellites. Since the full width at half maximum (FWHM) of the main peak is 150 Hz, the $T_2^*$ value obtained as the inverse of the FWHM would be 2.2 ms. Quite generally, however, magnetic inhomogeneity as well as phase incoherence causes the line broadening (which is called "inhomogeneous broadening"). Thus, the $T_2^*$ value of 2.2 ms evaluated from the FWHM should be regarded as the lower limit of $T_2$. Intrinsic $T_2$ value in the absence of magnetic inhomogeneity is given from a spin-echo method, as is shown below.

At about 0.5 kHz higher and lower frequencies of the main peak, small satellites are observed. As was clarified by the detailed study on both "29 silicon" and "natural silicon" NMR spectra, [4] we can safely say that the satellites are the "remnant" of the Pake's doublet. In other words, a very small number of $^{29}$Si -$^{29}$Si dimers cause the doublet via the direct nuclear-dipolar interaction.

Figure 2(a) shows the multiple spin-echoes obtained by the APCP pulse sequence. As is shown in Fig. 2(b), the time dependence of the spin-echo intensity greatly deviates from a single Gaussian or Lorentzian function, but is found to be well-reproduced by a bi-exponential function such that

$A\exp(-t/T_2^S) + (1-A)\exp(-t/T_2^L)$,

where $T_2^S = 15\pm5$ ms, $T_2^L = 200\pm20$ ms and $A = 0.65\pm015$.
As expected, even the $T_2^S$ is about ten times as great as the $T_2^*$ estimated from the FWHM of the spectrum.

It should be noted that the $T_2^L/T_2^S$ value is of the order of 10. It is tempting to think that, taking into account the presence of the Pake's doublet, the fast-decaying component characterized by the $T_2^S$ would come from the direct nuclear-dipolar interaction present between the nearest neighbor $^{29}$Si and $^{29}$Si. In this case, according to the ratio of the intensity of the satellites over the main-peak intensity, the fraction $A$ should be of the order of $10^{-2}$. Clearly, this is much smaller than the value experimentally obtained. At present, we have no way to clarify the great deviation from a single Gaussian or Lorentzian function.

One might argue that only the value of $T_2^S$ should be important to discuss whether pure silicon is promising for a quantum computer or not. However, this is not correct, because the nuclear spins do retain their coherence in the $T_2^L$ region. Note that the spin-echoes can be observed even in the $T_2^L$ region without averaging.

In conclusion, for pure and carrier-less "natural silicon", we performed pulsed NMR experiments at room temperature. The value of FWHM of the observed $^{29}$Si NMR line shape gives $T_2^*$ of 2.2 ms. From the time dependence of the FID intensity after saturation, the $T_1$ value is estimated to be of the order of $10^4$ s. To solve the waiting-time problem coming from the extremely long $T_1$, we controlled the room temperature within the error of 0.5 K. By employing the APCP pulse sequence, we have succeeded in observing a series of spin-echoes of which the time dependence is characterized by $T_2$. To our knowledge, this is the first report on precise measurement of $T_2$ at room temperature for pure and carrier-less "natural silicon". The decoherence process is found to deviate from a single Gaussian or Lorentzian function but to be well-characterized by $T_2^S = 15 \pm 5$ ms and $T_2^L = 200 \pm 20$ ms. These $T_2$ values experimentally prove that the pure and carrier-less "natural silicon" is a promising material for nuclear-spin computing.

The authors are grateful to Kohei M. Itoh and Eeisuke Abe for providing the sample and stimulating discussions. This work was supported by the Core Research for Evolutional Science and Technology (CREST) Program under the Japan Science and Technology Corporation (JST) initiative.

After we finished our work, we were informed that Dementyev and his colleagues reported the similar results[9] as ours. However, our data obtained for $t < 360$ ms [Fig. 2(b)] completely includes their data at room temperature only within $t < 80$ ms.

Fig. 1. $^{29}$Si-NMR spectrum of pure (99.99999 %, 7N) and carrier-less (p-type with doping level $1\times10^{15}$ cm$^{-3}$) powdered silicon in which $^{29}$Si nuclei are naturally abundant (4.7%). The horizontal axis is offset by Larmor frequency (58.430MHz). The spectrum consists of a main peak (FWHM $\sim$ 150 Hz) and two satellites. The $T_2$* value obtained as the inverse of the FWHM is 2.2 ms. The satellites are due to the direct nuclear-dipolar interaction between the nearest-neighbor $^{29}$Si nuclei.

Fig. 2(a) $^{29}$Si nuclear-spin decoherence process directly observed by multiple spin-echoes for the pure and carrier-less naturally abundant silicon at room temperature. To observe the multiple spin-echoes, the alternating phase Carr-Purcell (APCP) pulse sequence (not shown in the figure for simplicity) is employed.

Fig. 2(b) Time dependence of the even-number spin-echo intensities obtained by the APCP pulse sequence. The time dependence is characterized by the $^{29}$Si nuclear-spin decoherence time $T_2$. The decoherence process is found to deviate from a single Gaussian (broken line) or Lorentzian (straight line, not shown) function but to be well-reproduced by a bi-exponential function (solid line) with the shorter $T_2^S = 15\pm5$ ms and the longer $T_2^L = 200\pm20$ ms.

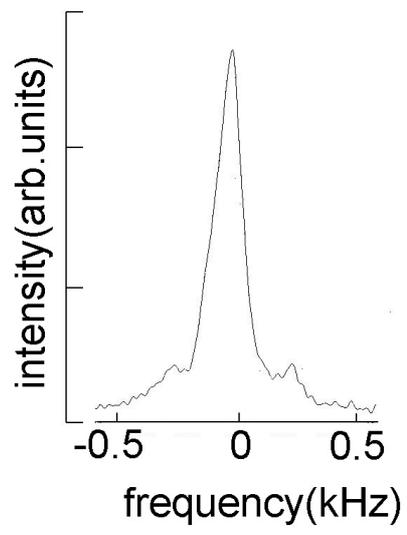

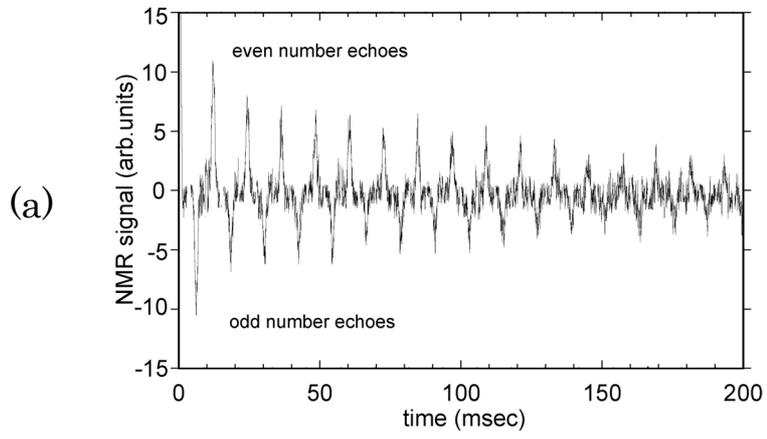

(a)

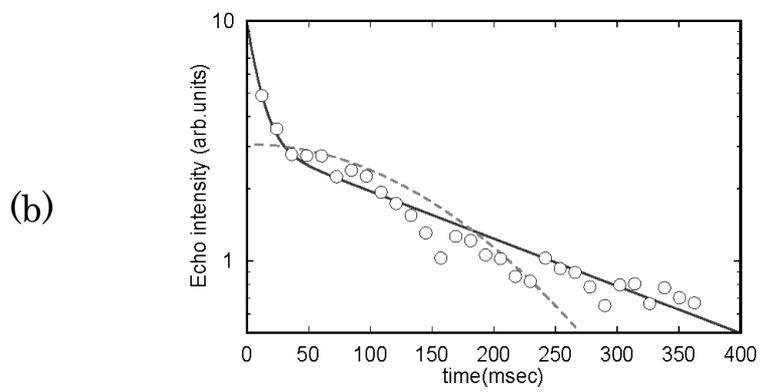

(b)